\documentclass[12pt,reqno]{amsart}
\usepackage{amsfonts}
\usepackage{amssymb}
\usepackage{amsmath}
\usepackage{amsthm}
\usepackage{txfonts}
\usepackage{enumerate}
\usepackage{fullpage}
\usepackage{hangcaption}
\usepackage{fancyhdr}

\makeatletter\@addtoreset{equation}{section}\makeatother

\allowdisplaybreaks
\newtheorem{thm}{Theorem}[section]

\newtheorem{defn}[thm]{Definition}
\newtheorem{kor}[thm]{Corollary}
\newtheorem{lem}[thm]{Lemma}
\newtheorem{rem}[thm]{Remark}

\newtheorem{ass}[thm]{Assumption}

\begin{document}
\title{Feynman integrals as Hida distributions: the case of non-perturbative
potentials\bigskip\\ \small\small\small {Dedicated to Jean-Michel Bismut as a small token of appreciation}}



\author{Martin Grothaus, Ludwig Streit, Anna Vogel}
\address{Martin Grothaus, Mathematics Department, University of
Kaiserslautern \\
P.O.Box 3049, 67653 Kaiserslautern,
\textrm{\texttt{Email: grothaus@mathematik.uni-kl.de}\\ \newline\texttt{URL:
http://www.mathematik.uni-kl.de/$\sim$grothaus/, }} \newline
Ludwig Streit, CCM, Universidade da Madeira \\
9000-390 Funchal, Protugal; BiBoS-Bielefeld University, 33615 Bielefeld,
Germany \\\textrm{\texttt{Email: streit@Physik.Uni-Bielefeld.de}} \\\newline
Anna Vogel, Mathematics Department, University of Kaiserslautern \\
P.O.Box 3049, 67653 Kaiserslautern, Germany;
\textrm{\texttt{Email: vogel@mathematik.uni-kl.de}}}
\date{\today}
\maketitle

\section{Introduction}

Feynman "integrals", such as%
\begin{equation*}
J=\int d^{\infty }x\exp \left( i\int_{0}^{t}\left( T(\dot{x}(s))-V\left(
x(s)\right) \right) ds\right) f(x(\cdot ))
\end{equation*}%
are commonplace in physics and meaningless mathematically as they stand. Within white noise analysis \cite{BeKo88,BCB02,Hi80,HKPS93,KLPSW96,Kuo96,L06,P91,SS02} the concept of integral has a natural
extension in the dual pairing of generalized and test functions and allows
for the construction of generalized functions (the "Feynman integrands") for
various classes of interaction potentials $V$, see e.g.~\cite{FPS91,FOS05,GKSS97,HKPS93,KS92,KWS97,SS02}, all of
them by perturbative methods. This work extends this framework to the case where these fail, using
complex scaling as in \cite{D80}, see also \cite{C61}.

In Section 2 we characterize Hida distributions. In Section 3 the $U$%
-functional is constructed, see Theorem \ref{ufunct}. We prove in Section 4
that we obtain a solution of the Schroedinger equation, see Theorem \ref%
{thmschreq}. The strategy for a general construction of the Feynman integrand is provided in Section 5. Examples are given in Section 6.

\section{White Noise Analysis}

\label{ss31}

The white noise measure $\mu $ on Schwartz distribution space arises from
the characteristic function%
\begin{equation*}
C(f):=\exp \Big(-{\textstyle\frac{1}{2}}\Vert f\Vert _{2}^{2}\Big),\quad
f\in S({\mathbb{R}}),
\end{equation*}%
via Minlos' theorem, see e.g.  \cite{BeKo88,Hi80,HKPS93}:
\begin{equation*}
C(f)=\int_{S^{\prime }}\exp \big(i\langle \omega ,f\rangle \big)\,d{\mu
(\omega )}.
\end{equation*}%
Here $\langle \cdot ,\cdot \rangle $ denotes the dual pairing of $S^{\prime }({%
\mathbb{R}})$ and $S({\mathbb{R}})$. We define the space
\begin{equation*}
\big( L^{2}\big) :=L^{2}(S^{\prime }({\mathbb{R}}),\mathcal{B},\mu ).
\end{equation*}%
In the sense of an $L^{2}$-limit to indicator functions $\mathbf{1}_{[0,t)},t>0,$
a version of Wiener's Brownian motion is given by:
\begin{equation*}
B(t,\omega ):=\langle \omega ,\mathbf{1}_{[0,t)}\rangle =\int_{0}^{t}\omega
(s)~ds,~~t>0.
\end{equation*}%
One then constructs a Gel'fand triple:
\begin{equation*}
(S)\subset L^{2}(\mu )\subset (S)^{\prime }
\end{equation*}%
of Hida test functions and distributions, see e.g.~\cite{HKPS93}. We introduce the $T$-transform of $\Phi \in
(S)^{\prime }$ by
\begin{equation*}
(T\Phi )(g):=\big\langle\!\big\langle\Phi ,\exp \big(i\langle \cdot
,g\rangle \big)\big\rangle\!\big\rangle\!~\!,\quad  g\in S(\mathbb{R}),
\end{equation*}
where $\langle \!\langle \cdot ,\cdot \rangle \!\rangle $ denotes the
bilinear dual pairing between $(S)^{\prime }$ and $(S)$. Expectation extends
to Hida distributions $\Phi $ by
\begin{equation*}
E_{\mu }(\Phi ):=\langle \!\langle \Phi ,1\rangle \!\rangle .
\end{equation*}

\begin{defn}
\label{defufunc} A function $F:S(\mathbb{R})\rightarrow \mathbb{C}$ is
called $U$-functional if

\begin{description}
\item[(i)] $F$ is "ray-analytic": for all $g,h\in S(\mathbb{R})$ the mapping
\begin{equation*}
\mathbb{R}\ni y\mapsto F(g+yh)\in \mathbb{C}
\end{equation*}%
has an analytic continuation to $\mathbb{C}$ as an entire function.

\item[(ii)] F is uniformly bounded of order 2, i.e., there exist some
constants $0<K,D<\infty$ and a continuous norm $\|\cdot\|$ on $S(\mathbb{R}%
)$ such that for all $w\in\mathbb{C}$, $g\in S(\mathbb{R})$
\begin{align*}
|F(wg)|\leq K\exp(D|w|^2\|g\|^2).
\end{align*}
\end{description}
\end{defn}

\begin{thm}
\label{charac} The following statements are equivalent:

\begin{description}
\item[(i)] $F:S(\mathbb{R})\to \mathbb{C}$ is a $U$-functional.

\item[(ii)] $F$ is the $T$-transform of a unique Hida distribution $\Phi \in
(S)^{\prime }$.
\end{description}
\end{thm}

For the proof and more see e.g. \cite{HKPS93}.

\section{Hida distributions as candidates for Feynman Integrands}

\label{constru} In this section we construct a Hida distribution as a
candidate for the Feynman integrand. First we list which properties
potentials must fulfill.

\begin{ass}
\label{ASSI}
For $\mathcal{O}\subset \mathbb{R}$ open, where $\mathbb{R}\setminus
\mathcal{O}$ is a set of Lebesgue measure zero, we define the set $\mathcal{D}%
\subset \mathbb{C}$ by
\begin{equation*}
\mathcal{D}:=\Big\{x+\sqrt{i}y~\Big|~x\in \mathcal{O}~\text{and}~y\in
\mathbb{R}\Big\},
\end{equation*}
and consider analytic functions ${V}_{0}:\mathcal{D}%
\rightarrow \mathbb{C}$ and $f:\mathbb{C}\rightarrow \mathbb{C}$. Let $0\leq t\leq T<\infty$. We require that
there exists an $0<\varepsilon <1$ and a 
function $I:\mathcal{D}\rightarrow \mathbb{R}$ such that its restriction to ${\mathcal{O}}$ is measurable and locally bounded and
\begin{equation}  \label{intab}
E\Bigg[\Bigg|\exp \Bigg(-i\int_{0}^{t}{V}_{0}\Big(z+\sqrt{i}B_{s}\Big)%
ds\Bigg){f}\Big(z+\sqrt{i}B_{t}\Big)\Bigg|\exp \Bigg(\frac{\varepsilon
\Vert B\Vert _{{\sup ,T}}^{2}}{2}\Bigg)\Bigg]\leq I(z),\quad z\in \mathcal{D},
\end{equation}%
uniformly in $0\leq t\leq T$.
Here $E$ denotes the expectation w.r.t.~a Brownian motion $B$ starting at 0. $\Vert \cdot
\Vert _{{\sup ,T}}$ denotes the supremum norm over $[0,T]$.
\end{ass}
We shall consider time-dependent potentials of the form
\begin{align}
{V}_{\dot{g}}:[0,T]\times \mathcal{D}& \rightarrow \mathbb{C}  \notag \\
(t,z)& \mapsto {V}_{0}(z)+\dot{g}(t)z  \label{pot}
\end{align}%
for $g\in S(\mathbb{R}).$

\begin{rem}
One can show that (\ref{intab}) implies that
\begin{equation*}
E\Bigg[\exp \Bigg(-i\int_{0}^{t-t_0}{V}_{\dot{g}}\Big(t-s,z+\sqrt{i}B_{s}%
\Big)ds\Bigg){f}\Big(z+\sqrt{i}B_{t-t_0}\Big)\Bigg],
\end{equation*}%
is well-defined for all $g\in S(\mathbb{R})$, $0\leq t_0\leq t\leq T$ and $z\in
\mathcal{D}$.
\end{rem}

\begin{thm}
\label{ufunct} Let $0<T<\infty$ and $%
\varphi:\mathbb R\to\mathbb R $ be Borel measurable, bounded with compact support. Moreover we assume that $V_0$ and $f$ fulfill Assumption \ref{ASSI}. Then for all $%
0\leq t_0\leq t\leq T$, the mapping
\begin{align}
F_{\varphi ,t,{f},t_0}&:S(\mathbb{R}) \rightarrow \mathbb{C}  \notag \\
g~& \mapsto \exp \Bigg(-\frac{1}{2}\int_{[t_0,t]^{c}}g^{2}(s)~ds\Bigg)\int_{%
\mathbb{R}}\exp (-ig(t_0)x)\varphi (x)\bigg(G(g,t,t_0)\exp (ig(t)\cdot ){f}\bigg)%
(x)~dx  \label{deFI2}
\end{align}%
is a $U$-functional where for $x\in \mathcal{O}$
\begin{multline}
\bigg(G(g,t,t_0)\exp (ig(t)\cdot ){f}\bigg)(x):=E\Bigg[\exp \Bigg(-i\int_{0}^{t-t_0}%
{V}_{\dot{g}}\Big(t-s,x+\sqrt{i}B_{s}\Big)~ds\Bigg) \\
\times \exp \Big(ig(t)\Big(x+\sqrt{i}B_{t-t_0}\Big)\Big){f}\Big(x+\sqrt{i}%
B_{t-t_0}\Big)\Bigg].  \label{G}
\end{multline}
\end{thm}

\begin{proof}
$F_{\varphi ,t,{f},t_0}$ is well-defined: (\ref{G}) is finite because of (\ref%
{ASSI}), and the integral in (\ref{deFI2}) exists since $\varphi $ is bounded with
compact support.

To show that $F_{\varphi ,t,{f},t_0}$ is a $U$-functional we must verify two
properties, see Definition \ref{defufunc}.

First $F_{\varphi ,t,{f},t_0}$ must have a "ray-analytic" continuation to $\mathbb{%
C}$ as an entire function. I.e., for all $g,h\in S(\mathbb{R})$ the mapping
\begin{equation*}
\mathbb{R}\ni y\mapsto F_{\varphi ,t,{f},t_0}(g+yh)\in \mathbb{C}
\end{equation*}%
has an entire extension to $\mathbb{C}$.

We note first that this is true for the expression
\begin{multline}
u(y):=\exp \Bigg(-i\int_{0}^{t-t_0}{V}_{\dot{g}+y \dot{h}}\Big(%
t-s,x+\sqrt{i}B_{s}\Big)~ds\Bigg) \\
\times \exp \Big(i\left( g+yh\right) (t)\Big(x+\sqrt{i}B_{t-t_0}\Big)\Big){%
f}\Big(x+\sqrt{i}B_{t-t_0}\Big)
\end{multline}%
inside the expectation in (\ref{G}). Hence the integral of $u$ over any closed
curve in $\mathbb{C}$ is zero. By Lebesgue dominated convergence the
expectation $E[u(w)]$ is continuous in $w$. With Fubini%
\begin{equation*}
\oint E\left[ u\left( w\right) \right] dw=E\left[ \oint u\left( w\right)
dw\right] =0,
\end{equation*}%
for all closed paths, hence by Morera $E(u(w))$ is entire. This extends to (%
\ref{deFI2}) since $\varphi $ is bounded with compact support. Thus
\begin{equation*}
\mathbb C\ni w\mapsto F_{\varphi ,t,{f},t_0}(g+wh)\in\mathbb C
\end{equation*}%
is entire for all $0\leq t_0\leq t\leq T$ and all $g,h\in S(\mathbb{R})$.

Verification is straightforward that $F_{\varphi ,t,{f},t_0}$ is of 2nd order
exponential growth, $F_{\varphi ,t,{f},t_0}$ is a U-functional.
\end{proof}

One can show the same result by choosing the delta distribution $\delta
_{x},~x\in \mathcal{O}$, instead of a test function $\varphi $:

\begin{kor}
\label{ufunct2} Let $V_0$ and $f$ fulfill Assumption \ref{ASSI} and let $%
x\in \mathcal{O}$. Then for all $0\leq t_0\leq t\leq T$ the mapping
\begin{align*}
F_{\delta_x,t,{f},t_0}:S(\mathbb{R})&\to\mathbb{C}  \notag \\
g~&\mapsto \exp\Bigg(-\frac{1}{2}\int_{[t_0,t]^{c}}g^{2}(s)~ds\Bigg)\exp
(-ig(t_0)x)\Big(G(g,t,t_0)\exp (ig(t)\cdot)\Big){f}(x)
\end{align*}
is a $U$-functional, where $\Big(G(g,t,t_0)\exp (ig(t)\cdot)\Big){f}(x)$ is
defined as in Theorem \ref{ufunct}.
\end{kor}

\section{Solution to time-dependent Schr\"{o}dinger equation}

\begin{ass}\label{ASSII}
 Let  ${V}_0:\mathcal{D}\to\mathbb C$ and ${f}:\mathbb C\to\mathbb C$ such that Assumption \ref{ASSI} is fulfilled and
${V}_{\dot{g}},~g\in S(\mathbb R)$, as in (\ref{pot}).
\begin{description}
\item[(i)] For all $u,v,r,l\in[0,T]$ and all $z\in{\mathcal D}$ we require that
\begin{align}
		E^1\Bigg[\Bigg|&\exp \Bigg(-i\int_{0}^{u}{V}_{\dot{g}}\Big(v-s,z+\sqrt{i}B^1_{s}\Big)~
ds\Bigg)\notag\\
\times E^2\Bigg[&\exp\Bigg(-i\int_0^{r}{V}_{\dot{g}}\Big(l-s,z+\sqrt{i}B^1_{u}+\sqrt{i}B^2_{s}\Big)~ds\Bigg){f}\Big(z+\sqrt{i}B^1_{u}+\sqrt{i}B^2_{r}\Big)\Bigg]\Bigg|\Bigg]<\infty.\label{abschJ}
\end{align}
\item[(ii)] For all $z\in{\mathcal D}$, $0\leq t_0\leq t\leq T$ and some $0<\varepsilon\leq T$ the functions
	\begin{align}
		\omega\mapsto 						\sup_{0\leq h\leq\varepsilon}\Bigg|&\Bigg({V}_{\dot{g}}\Big(t,z+\sqrt{i}B_{h}(\omega)\Big)+\int_0^h \frac{\partial}{\partial t} {V}_{\dot{g}}\Big(t+h-s,z+\sqrt{i}B_{s}(\omega)\Big)~ds\Bigg)\notag\\&\times\exp\Bigg(-i\int_0^{h}{V}_{\dot{g}}\Big(t+h-s,z+\sqrt{i}B_{s}(\omega)\Big)~ds\Bigg){f}\Big(z+\sqrt{i}B_{h}(\omega)\Big)\Bigg|\label{VvonUtt0}
		\end{align}
		and
			\begin{align}
		\omega\mapsto 					\sup_{h\in[0,T]}\Bigg|\Delta E^2\Bigg[&\exp\Bigg(-i\int_0^{t-t_0} {V}_{\dot{g}}\Big(t-s,z+\sqrt{i}B^1_{h}(\omega)+\sqrt{i}B^2_{s}\Big)~ds\Bigg)\notag\\&\times{f}\Big(z+\sqrt{i}B^1_{h}(\omega)+\sqrt{i}B^2_{t-t_0}\Big)\Bigg]\Bigg|\label{HvonUtt0}
		\end{align}
		are integrable.
\end{description} 
Here $B^1$ and $B^2$ are Brownian motions starting at 0 with corresponding expectations $E^1$ and $E^2$, respectively. Moreover $\Delta$ denotes $\frac{\partial^2}{\partial z^2}$ and $\frac{\partial}{\partial t}$ the derivative w.r.t.~the first variable. 
\end{ass}	
We define $H({\mathcal D})$ to be the set of holomorphic functions from ${\mathcal D}$ to $\mathbb C$.
As pointed out by H. Doss, see \cite{D80}, under specified assumptions (similar to Assumption \ref{ASSI} and Assumption \ref{ASSII} (ii)) there is a solution
$\psi:[0,T]\times {\mathcal D}\to \mathbb C$ to the time-independent Schr\"odinger
equation, i.e., for all $t\in[0,T]$ and $x\in\mathcal{O}$
\begin{align*}
 \begin{cases} {i\frac{\partial}{\partial
t}\psi(t,x)=-\frac{1}{2}\Delta\psi(t,x)+{V}_0(x)\psi(t,x)}\\
{~~~~~~~~~~\psi(0,x)=f(x)},
\end{cases}
\end{align*}
which is given by
\begin{align*}
        \psi(t,x)=E\Bigg[\exp\Bigg(-i\int_0^{t}{V}_0\Big(x+\sqrt{i}B_s\Big)ds\Bigg){f}\Big(x+\sqrt{i}B_{t}\Big)\Bigg].
\end{align*}
\begin{rem}\label{freecase}
	Let us consider the case of the free motion, i.e., $V_0\equiv 0$. We assume that $f:\mathcal D\to\mathbb C$ is an analytic function, such that $E\Big[{f}\Big(z+\sqrt{i}B_t\Big)\Big]$, $z\in\mathcal D$, $0\leq t\leq T$, exists and is uniformly bounded on $[0,T]$. Moreover let
			\begin{align*}
		\omega\mapsto 						\sup_{h\in[0,T]}\Big|\Delta {f}\Big(z+\sqrt{i}B_{h}(\omega)\Big)~\!\Big|
		\end{align*}
	 be integrable, then
	\begin{align*}
			\frac{\partial}{\partial t}E\Big[{f}\Big(x+\sqrt{i}B_t\Big)\Big]=-i\frac{1}{2}\Delta E\Big[{f}\Big(x+\sqrt{i}B_t\Big)\Big],
	\end{align*}
	for $x\in\mathcal O$, $0\leq t\leq T$.
\end{rem} 
For our purpose a generalization to the time-dependent case
\begin{equation}
\begin{cases}
{i\frac{\partial }{\partial t}\big(U(t,t_0)f\big)(x)=\big(H(t)U(t,t_0)f\big)(x)} \\
{~~~~~~~~~~\big(U(t_0,t_0)f\big)(x)=f(x)},%
\end{cases}
\quad \quad x\in {\mathcal O},~0\leq t_0\leq t\leq T,\label{schreq}
\end{equation}%
where $H(t):=-\frac{1}{2}\Delta +{V}_{\dot{g}}(t,\cdot)$ for $g\in S(\mathbb{R})$ and $0\leq t\leq T$, is necessary.
In the following we show that the operator $U(t,t_0):D(t,t_0)\subset H({\mathcal D})\to H({\mathcal D})$, $0\leq t_0\leq t\leq T$, given by
\begin{align} U(t,t_0){f}(z):=E\Bigg[\exp\Bigg(-i\int_0^{t-t_0}{V}_{\dot{g}}\Big(t-s,z+\sqrt{i}B_{s}\Big)ds\Bigg){f}\Big(z+\sqrt{i}B_{t-t_0}\Big)\Bigg],\quad z\in\mathcal D,\label{utt0}
\end{align}
provides us with a solution to (\ref{schreq}). Here by $D(t,t_0)$ we denote the set of functions ${f}\in H({\mathcal D})$ such that the expectation in (\ref{utt0}) is a well-defined object in $H(\mathcal{D})$.
\begin{lem}\label{lemutt0}
Let $V_0$ and $f$ fulfill the Assumptions \ref{ASSI} and \ref{ASSII} then the operator $U(t,t_0)$, $0\leq t_0\leq t\leq T$, as in (\ref{utt0}), maps from $D(t,t_0)$ to $H({\mathcal D})$. Moreover $U(r,t_0){f}\in D(t,r)$ and one gets that
\begin{align*}
	U(t,t_0){f}(z)=U(t,r)(U(r,t_0){f})(z),
\end{align*}
for all $0\leq t_0\leq r\leq t\leq T$ and $z\in {\mathcal D}$.
\end{lem}
\begin{proof}
The property that $U(t,t_0)$, $0\leq t_0\leq t\leq T$, as in (\ref{utt0}), maps from $D(t,t_0)$ to $H({\mathcal D})$ follows by using Morera and Assumption \ref{ASSI}. The fact that $U(r,t_0){f}\in D(t,r)$ follows by Assumption \ref{ASSII} (i).
Let $0\leq t_0\leq r\leq t\leq T$ and $z\in D$, then
one gets with the Markov property and the time-reversibility of Brownian motion that
\begin{multline}  U(t,t_0)f(z)=E\Bigg[\exp\Bigg(-i\int_0^{t-t_0}{V}_{\dot{g}}\Big(t-s,z+\sqrt{i}B_{s}\Big)ds\Bigg){f}\Big(z+\sqrt{i}B_{t-t_0}\Big)\Bigg]\\ 
     =E\Bigg[\exp\Bigg(-i\int_0^{t-r}{V}_{\dot{g}}\Big(t-s,z+\sqrt{i}B_{s}\Big)ds\Bigg)\\
 \times\exp\Bigg(-i\int_{t-r}^{t-r+r-t_0}{V}_{\dot{g}}\Big(t-s,z+\sqrt{i}B_{s}\Big)ds\Bigg){f}\Big(z+\sqrt{i}B_{t-t_0}\Big)\Bigg]\\
     =E\Bigg[\exp\Bigg(-i\int_0^{t-r}{V}_{\dot{g}}\Big(t-s,z+\sqrt{i}B_{s}\Big)ds\Bigg)\\
     \times \exp\Bigg(-i\int_{0}^{r-t_0}{V}_{\dot{g}}\Big(r-s+t-r,z+\sqrt{i}B_{s+t-r}\Big)ds\Bigg){f}\Big(z+\sqrt{i}B_{t-r+r-t_0}\Big)\Bigg]\\
     =E^1\Bigg[\exp\Bigg(-i\int_0^{t-r}{V}_{\dot{g}}\Big(t-s,z+\sqrt{i}B^1_{s}\Big)ds\Bigg)\\  \times E^2\Bigg[\exp\Bigg(-i\int_{0}^{r-t_0}{V}_{\dot{g}}\Big(r-s,z+\sqrt{i}B^1_{t-r}+\sqrt{i}B^2_{s}\Big)ds\Bigg){f}\Big(z+\sqrt{i}B^1_{t-r}+\sqrt{i}B^2_{r-t_0}\Big)\Bigg]\Bigg]\\
     =U(t,r)\big(U(r,t_0)f\big)(z).
\end{multline}
\end{proof}
One can show that by $U(t,t_0)$, $0\leq t_0\leq t\leq T$, a pointwise-defined (unbounded) evolution system is given.
\begin{thm}\label{thmschreq}
	Let $0<T<\infty$, $V_0$, $V_{\dot g}$, $g\in
S(\mathbb{R})$, as in (\ref{pot}), and $f$ such that Assumption \ref{ASSI} and \ref{ASSII} are fulfilled.
	Then $U(t,t_0){f}(x)$, $0\leq t_0< t\leq T$, $x\in\mathcal{O}$, given in (\ref{utt0}) solves the Schr\"odinger equation (\ref{schreq}).
\end{thm}
\begin{proof}
Let $0\leq t_0<t\leq T$, $x\in {\mathcal O}$ and $g\in
S(\mathbb{R})$.
If we have a look at the difference quotient from the right side, we get with Lemma \ref{lemutt0} that
\begin{align*}
	\frac{\partial}{\partial t}^+ U(t,t_0){f}(x)&=\lim_{h\searrow 0}\frac{U(t+h,t_0)-U(t,t_0)}{h}{f}(x)\notag\\&=\lim_{h\searrow 0}\frac{U(t+h,t)-U(t,t)}{h}U(t,t_0){f}(x).
\end{align*}
Hence it is left to show that
\begin{align*}
 \lim_{h\searrow 0}\frac{U(t+h,t)k(x)-U(t,t)k(x)}{h}=H(t)k(x),
\end{align*}
for $k=U(t,t_0)f$. 
Note that
\begin{multline}
	\lim_{h\searrow 0}\frac{1}{h}E\Bigg[\exp\Bigg(\int_0^{t+h-t} {V}_{\dot{g}}\Big(t+h-s,x+\sqrt{i}B_s\Big)ds\Bigg)k\Big(x+\sqrt{i}B_h\Big)-k\Big(x+\sqrt{i}B_0\Big)\Bigg]\\
	=\lim_{h\searrow 0}E\Bigg[ \frac{1}{h}\exp\Bigg(\int_0^h {V}_{\dot{g}}\Big(t+h-s,x+\sqrt{i}B_s\Big)ds\Bigg)k\Big(x+\sqrt{i}B_h\Big)-\frac{1}{h}k\Big(x+\sqrt{i}B_h\Big)\Bigg]\\+\lim_{h\searrow 0}E\Bigg[\frac{1}{h}k\Big(x+\sqrt{i}B_h\Big)-\frac{1}{h}k\Big(x+\sqrt{i}B_0\Big)\Bigg].
\end{multline}
The integrand of the first summand yields
\begin{align*}
	\lim_{h\searrow 0}&\frac{1}{h}\Bigg(\exp\Bigg(\int_0^h {V}_{\dot{g}}\Big(t+h-s,x+\sqrt{i}B_s\Big)ds\Bigg)-1\Bigg)k\Big(x+\sqrt{i}B_h\Big)\\
 =&{V}_{\dot{g}}\Big(t,x+\sqrt{i}B_0\Big) k\Big(x+\sqrt{i}B_0\Big)={V}_{\dot{g}}(t,x) k(x).
\end{align*}
Hence by Assumption \ref{ASSII} (ii), the mean value theorem and Lebesgue dominated convergence
\begin{align*}
\lim_{h\searrow 0}E\Bigg[\frac{1}{h}\Bigg(\exp\Bigg(\int_0^h {V}_{\dot{g}}\Big(t+h-s,x+\sqrt{i}B_s\Big)ds\Bigg)-1\Bigg)k\Big(x+\sqrt{i}B_h)\Big)\Bigg]={V}_{\dot{g}}(t,x) k(x).
\end{align*}
Moreover we know by Remark \ref{freecase} and Assumption \ref{ASSII} (ii) that
$E\Big[k(x+\sqrt{i}B_t)\Big]$ solves the free Schr\"odinger equation, hence
\begin{align*}
\lim_{h\searrow 0}E\Big[ k\Big(x+\sqrt{i}B_h\Big)-k\Big(x+\sqrt{i}B_0\Big)\Big]=-i\Delta k(x).
\end{align*}
Similar with
\begin{align*}
	\frac{\partial}{\partial t}^- U(t,t_0){f}(x)&=\lim_{h\searrow 0}\frac{U(t-h,t_0)-U(t,t_0)}{h}{f}(x)\notag\\&=\lim_{h\searrow 0}\frac{U(t-h,t-h)-U(t,t-h)}{h}U(t-h,t_0){f}(x)
\end{align*}
one can show the same for the difference quotient from the left side.
\end{proof}
\section{General construction of the Feynman integrand}

\label{deffy} Of course one is interested in the Feynman integrand
$I_{V_0}$
for a general class of potentials $V_0:\mathcal{O} \to \mathbb{C}$, where $%
\mathbb{R}\setminus \mathcal{O}$ is of measure zero, having an analytic continuation to $\mathcal{D}$. I.e., we are interested in the Feynman integrand corresponding to the
Hamiltonian
\begin{align*}
H^{(0)}=-\frac{1}{2}\Delta+V_0(q),
\end{align*}
where $q$ is the position operator, i.e.,
\begin{align*}
H^{(0)}\varphi(x)=-\frac{1}{2}\Delta\varphi(x)+V_0(x)\varphi(x),~~\quad%
~x\in\mathcal{O},
\end{align*}
for suitable $\varphi:\mathbb R\to\mathbb R$
(see the introduction for a comprehensive list of references). In
all cases it turned out that for a test function $g\in
S(\mathbb{R})$ and $0\leq t_0 < t\leq T $ we have that
\begin{align}
(TI_{V_0})(g)=\exp\Bigg(-\frac{1}{2}\big\|g\mathbf{1}_{[t_0,t]^c}\big\|^2+ig(t)
x-ig(t_0) x_0\Bigg) K_{V_0}^{(\dot{g})}(x,t|x_0,t_0),\label{tiv}
\end{align}
where $K_{V_0}^{(\dot{g})}(x,t|x_0,t_0)$ denotes the Green's
function corresponding to the potential $V_{\dot{g}}$ (see \cite{GV08} for a justification of (\ref{tiv}) under natural assumptions on $I_{V_0}$). This leads
us to the following definition (see e.g. \cite{FPS91}).

\begin{defn}
\label{defFI1} Let $V_0:\mathcal{D}\to\mathbb{C}$ be an analytic potential, $f:%
\mathbb{C}\to\mathbb{C}$ an analytic initial state, $V_{\dot g}$, $g\in
S(\mathbb{R})$, as in (\ref{pot}), and $%
\varphi:\mathbb R\to\mathbb R $, Borel measurable, bounded with compact support. Assume
that $V_0$, ${V}_{\dot g}$ and $f$ fulfill Assumption \ref{ASSI} and Assumption \ref{ASSII}.
Then by Theorem \ref{ufunct} one has that for all $0\leq t_0\leq t\leq T$, 
the function $F_{\varphi,t,{f},t_0}$ exists and forms a
$U$-functional. Moreover by Theorem \ref{thmschreq} it follows that for all $x\in\mathcal D$ and all $0\leq t_0\leq t\leq T$
\begin{align*}
U_{\dot{g}}(t,t_0)f(x)=E\Bigg[\exp\Bigg(-i\int_0^{t-t_0}{V}_{\dot g}%
\Big(s,x+\sqrt{i}B_{s}\Big)ds\Bigg) {f}\Big(x+\sqrt{i}B_{t-t_0}%
\Big)\Bigg]
\end{align*}
exists and solves the Schr\"odinger equation (\ref{schreq})
corresponding to the Hamiltonian\newline
$H(t)=-\frac{1}{2}\Delta+V_0(q)+\dot{g}(t) q$ for all $g\in
S(\mathbb{R})$. Then by Theorem \ref{charac} we define the Feynman integrand
\begin{align*}
I_{V_0,\varphi,f}:=T^{-1} F_{\varphi,t,{f},t_0}\in (S)^{\prime}.
\end{align*}
\end{defn}

\begin{defn}
\label{defFI2} Again let $V_0:\mathcal{D}\to\mathbb{C}$ be an analytic potential, $f:%
\mathbb{C}\to\mathbb{C}$ an analytic initial state, $V_{\dot g}$, $g\in
S(\mathbb{R})$, as in (\ref{pot}) and $x\in\mathcal{O}$. Analogously with Theorem \ref{charac}, Corollary \ref{ufunct2} and
Theorem \ref{thmschreq} we define the Feynman integrand
\begin{align*}
I_{V_0,\delta_x,f}:=T^{-1}F_{\delta_x,t,{f},t_0}\in (S)^{\prime}.
\end{align*}
\end{defn}
\begin{rem}
Note that the Green's function $K_{V_0}^{(\dot{g})}(x,t|x_0,t_0)$, if it exists, is the integral kernel of the operator $U_{\dot{g}%
}(t,t_0)$.
\end{rem}

\section{Examples}

To show the existence of the Feynman integrand for concrete
examples one only has to verify Assumption \ref{ASSI} and
\ref{ASSII}. In this section we look at analytic potentials $V_0$ which
are already considered in \cite{D80}. First we introduce the
set of initial states $f$. For $m\in\mathbb N$ we choose the function
\begin{align}
{f}_m:\mathbb{C}&\to\mathbb{C} \notag\\
z&\mapsto (2^mm!)^{-\frac{1}{2}}(-1)^m\pi^{-\frac{1}{4}}e^{\frac{1}{2}z^2}%
\Bigg(\frac{\partial}{\partial z}\Bigg)^m e^{-z^2}.
\label{hermerw}
\end{align}
Note that the set of functions given by the restrictions of $f_m,~m\in\mathbb{N}$, to $\mathbb R$ are the Hermite
functions,
whose span is a dense subset of
$L^2(\mathbb{R})$.
\begin{lem}
\label{lembm} Let $k:\mathbb{R}_0^+\to \mathbb{R}_0^+$ be a measurable
function and $B$ a real-valued Brownian motion,
then
\begin{align*}
E\Big[k(\|B\|_{\sup,T})\Big]\leq 2\Bigg(\frac{2}{\pi T}\Bigg)%
^{1/2}\int_0^\infty k(u)e^{-\frac{u^2}{2T}}du.
\end{align*}
\end{lem}
For the proof see \cite[Sec.1, Lem.1]{D80}.
\begin{lem}
\label{lemherm} Let $f_m$, $m\in\mathbb{N}$, be as in (\ref{hermerw}). Then for all $l\in\mathbb{N}%
_0$ and $\varepsilon>0$ there exists a locally bounded measurable function $%
c_{m,l}:\mathbb{C}\to\mathbb{R}^+$ such that
\begin{align*}
\Big|{f}^{(l)}_m\Big(z+\sqrt{i}y\Big)\Big|\leq c_{m,l}(z)|y|^{m+l}\exp%
\Bigg(\Bigg(\frac{1}{2}+\frac{1}{\sqrt{2}\varepsilon}\Bigg)|z|^2\Bigg)\exp\bigg( \frac{%
\varepsilon}{2}|y|^2\bigg)\quad\mbox{for all}~ z\in\mathbb{C},y\in\mathbb{R},
\end{align*}
where ${f}^{(l)}_m$ denotes the $l$-th derivative of $f$.
\end{lem}

\subsection{The Feynman integrand for polynomial potentials}

Here for $n\in\mathbb{N}_0$ we have a look at the potential
\begin{align}
{V}_0:\mathbb{C}&\to\mathbb{C}  \notag \\
z&\mapsto (-1)^{n+1}a_{4n+2}z^{4n+2}+\sum_{j=1}^{4n+1}a_jz^j,
\label{holpoly}
\end{align}
for $a_0,.., a_{4n+1}\in\mathbb{C}$ and $a_{4n+2}>0$.
If we have a look at the function
\begin{align*}
y\mapsto -i{V}_{\dot g}\Big(t,x+\sqrt{i}y\Big)
\end{align*}
for $g\in S(\mathbb R)$, $x\in\mathbb{C}$ and $t\in[0,T]$, then it is easy to see that
the term of highest order of the real part is given by
$-a_{4n+2}y^{4n+2}$. So it follows that for all compact sets
$K\subset\mathbb{C}$ there exists a constant $C_K>0$ such that
\begin{align}
\sup_{z\in K}\sup_{t\in[0,T]}\sup_{y\in\mathbb{R}}~\bigg|\exp\Big(%
|g(t)|~(|z|+|y|)-i{V}_0\Big(z+\sqrt{i}y\Big)\Big)\bigg|\leq C_K.
\label{komp}
\end{align}
Hence the function
\begin{align}
\omega\mapsto\exp\Bigg(-i\int_0^{t}{V}_{\dot g}\Big(s,z+\sqrt{i%
}B_s(\omega)\Big)ds\Bigg)  \label{bv}
\end{align}
is bounded uniformly in $0\leq t\leq T$ and locally uniformly in
$z\in\mathbb{C}$.

\begin{thm}
\label{polythm} Let $0<T<\infty$, $V_0$ as in (\ref{holpoly}) and $f_m$, $m\in%
\mathbb{N}$, as in (\ref{hermerw}). Then it is possible to define the
corresponding Feynman integrand $I_{V_0,\varphi,f_m}$, $\varphi$ Borel measurable, bounded with compact support, and $I_{V_0,\delta_x,f_m}$, $x\in\mathbb R$,
as in Definition \ref{defFI1} and Definition \ref{defFI2},
respectively.
\end{thm}

\begin{proof}
As discussed above ${V}_0$ and
${f}_m$ are analytic. Moreover with Lemma \ref{lembm} and
\begin{align}
k_{z,l}:\mathbb{R}_0^+&\to\mathbb{R}_0^+  \notag \\
u&\mapsto c_{m,l}(z)u^{m+l}\exp%
\Bigg(\Bigg(\frac{1}{2}+\frac{1}{\sqrt{2}\varepsilon}\Bigg)|z|^2\Bigg)\exp\bigg( \frac{%
\varepsilon}{2}u^2\bigg),  \label{boundk}
\end{align}
$l\in\mathbb{N}_0$, we get that
\begin{align}
&E\Bigg[\exp\Bigg(\frac{\varepsilon \|B\|_{{\sup,T}}^2}{2}\Bigg) \Big|{%
f}_m\Big(z+\sqrt{i}B_t\Big)\Big|\Bigg] \leq E\Bigg[\exp\Bigg(\frac{%
\varepsilon \|B\|_{{\sup,T}}^2}{2}\Bigg)k_{z,0}(\|B\|_{{\sup,T}})\Bigg]%
<\infty,  \label{abschE}
\end{align}
for $0<\varepsilon<\frac{1}{2T}$, $z\in\mathbb{C}$ and $c_{m,l}$ as in Lemma %
\ref{lemherm}. If we multiply the integrand in (\ref{abschE}) with
the
bounded function in (\ref{bv}) we still have an integrable function for all $%
z\in\mathbb{C}$ and all $0\leq t\leq T$. So for showing Assumption
\ref{ASSI} one has to check whether there exists a
function $I:\mathbb{C}\to\mathbb{R}^+$ whose restriction to $\mathbb R$ is locally bounded and
measurable, such that relation (\ref%
{intab}) holds. It is easy to see that this is true for the
function
\begin{align*}
I:\mathbb{C}&\to\mathbb{R}^+ \\
z&\mapsto E\Bigg[\Bigg|\exp\Bigg(Re\Bigg(-i\int_0^{t}{V}_0\Big(z+%
\sqrt{i}B_{s}\Big)ds\Bigg)\Bigg)\Bigg|~ \exp\Bigg(\frac{\varepsilon \|B\|_{{%
\sup,T}}^2}{2}\Bigg)k_{z,0}(\|B\|_{{\sup,T}})\Bigg].
\end{align*}
The locally boundedness of the restriction of $I$ to $\mathbb R$ follows from (\ref{komp}) and the fact
that $c_{m,l}$ is locally bounded. Since $\mathcal{O}=\mathbb{R}$ one can choose an arbitrary $\varphi$, Borel measurable, bounded with compact support, to apply Theorem \ref{ufunct}.
Moreover if we omit the integration the assumptions of Corollary
\ref{ufunct2} are also fulfilled.

So it is only left to check whether Assumption \ref{ASSII} is
fulfilled. 
To show (\ref{abschJ}) again by the boundedness of \eqref{bv} one only has to show that
\begin{align*}
		&E^1\Bigg[\Bigg|E^2\Bigg[{f}_m\big(z+\sqrt{i}B^1_{t-r}+\sqrt{i}B^2_{r-t_0}\big)\Bigg]\Bigg|\Bigg]<\infty,\quad z\in\mathbb C,\quad 0\leq t_0\leq r\leq t\leq T.
\end{align*}
But this follows directly by Lemma \ref{lembm} and Lemma \ref{lemherm}. To show Assumption \ref{ASSII} (ii) note first that differentiation and integration in (\ref{HvonUtt0}) can be interchanged since the integrand is analytic
and its derivatives are integrable.
Since $V$ is polynomial, using the functions $k_{z,0}$, $k_{z,1}$ and $%
k_{z,2}$, see (\ref{boundk}), Lemma \ref{lembm} and Lemma \ref{lemherm} one can show a estimate similar to (\ref{abschE}) for (\ref{VvonUtt0}) and (\ref{HvonUtt0}), respectively. Hence they are integrable.
\end{proof}

\begin{rem}
For $n=0$ we are not dealing with the harmonic oscillator.
Nevertheless it is possible to handle a potential of the form
\begin{align*}
x\mapsto a_0+a_1x+a_2x^2,
\end{align*}
for $a_0,a_1\in\mathbb{C}$ and $a_2\in\mathbb{R}$ such that $|a_2|<\frac{1}{%
2T^2}$. In this case the function in (\ref{bv}) might be unbounded. So we have to
estimate the potential as in Lemma \ref{lemherm}, separately.
\end{rem}

\subsection{Non-perturbative accessible potentials}

In this section $\mathcal{O}=\mathbb{R}\setminus\{b\}$, $b\in\mathbb{R}$. We first consider analytic potentials of the form
\begin{align}
V_0:\mathcal{D}&\to\mathbb{C}  \notag \\
z&\mapsto \frac{a}{|z-b|^n},  \label{1durchh}
\end{align}
where $n\in\mathbb{N}$, $a\in\mathbb{C}$ and $b\in\mathbb{R}$.

\begin{lem}
\label{lem1x} Let $V_0$ be defined as in (\ref{1durchh}). Then $V_0$ is analytic on $\mathcal{D}$ and for all $z\in\mathcal{D}$, $z=x+\sqrt{i}y,~x\in\mathcal{O}, ~y\in\mathbb R$, and all $%
0\leq t\leq T $ we get that
\begin{align*}
\Big|V_0\Big(z+\sqrt{i}B_t\Big)\Big|=|a|~\Big|\exp\Big(-\frac{n}{2}\log\Big(%
\big(z-b+\sqrt{i}B_t\big)^2\Big)\Big|\leq|a|~\exp\Bigg(-\frac{n}{2}\log%
\Bigg(\frac{(x-b)^2}{2}\Bigg)\Bigg).
\end{align*}
\end{lem}
For the proof see \cite{D80}.
\begin{thm}
\label{Thm1x} Let $0<T<\infty$, $\mathcal{O}=\mathbb{R}\setminus\{b\}$, $%
V_0$ as in Lemma \ref{lem1x} and $f_m$, $m\in\mathbb{N}$, as in (\ref{hermerw}). Then it is possible to define the corresponding Feynman integrand $%
I_{V_0,\varphi,f_m}$, $\varphi:\mathbb{R}\setminus\{%
b\}\to\mathbb C$, Borel measurable, bounded with compact support and $I_{V_0,\delta_x,f_m}$, $x\in\mathbb{R}\setminus\{%
b\}$, as in Definition \ref{defFI1} and Definition
\ref{defFI2},
respectively.
\end{thm}

\begin{proof}
W.l.o.g.~we set $a=1$ and $b=0$. Then $\mathcal{O}=\mathbb{R}%
\setminus\{0\}$. So let $z\in\mathcal{D}$, $z=x+\sqrt{i}y,~x\in\mathcal{O}, ~y\in\mathbb R$, and $0\leq t\leq T$. Again we have to check Assumption
\ref{ASSI} and \ref{ASSII}. From Lemma \ref{lem1x} know that ${%
V}_0$ is analytic on $\mathcal{D}$.

Now we check whether relation (\ref{intab}) is true. From Lemma
\ref{lem1x} we know that
\begin{align*}
&\Bigg|\exp\Bigg(-i\int_0^{t}{V}_0\Big(z+\sqrt{i}B_{s}\Big)ds\Bigg)%
\exp\Bigg(\frac{\varepsilon \|B\|_{{\sup,T}}^2}{2}\Bigg){f}_m\Big(z+%
\sqrt{i}B_{t}\Big)\Bigg| \\
\leq &\exp \Bigg(t\exp\Bigg(-\frac{n}{2}\log\Bigg(\frac{x^2}{2}\Bigg)%
\Bigg)\Bigg)\exp\Bigg(\frac{\varepsilon \|B\|_{{\sup,T}}^2}{2}\Bigg)\Big|%
{f}_m\Big(z+\sqrt{i}B_{t}\Big)\Big|.
\end{align*}
So with
\begin{align}
k_{z,l}:\mathbb{R}_0^+&\to\mathbb{R}_0^+  \notag \\
u&\mapsto \exp \Bigg(T\exp\Bigg(-\frac{n}{2}\log\Bigg(\frac{x^2}{2}\Bigg)%
\Bigg)\Bigg)c_{m,l}(z)u^{m+l}\exp%
\Bigg(\Bigg(\frac{1}{2}+\frac{1}{\sqrt{2}\varepsilon}\Bigg)|z|^2\Bigg)\exp\bigg( \frac{%
\varepsilon}{2}u^2\bigg),  \label{boundk1}
\end{align}
$l\in\mathbb{N}_0$, Lemma \ref{lembm} and Lemma \ref{lemherm} we
get that
\begin{align}
&E\Bigg[\Bigg|\exp\Bigg(-i\int_0^{t}{V}_0\Big(z+\sqrt{i}B_{s}\Big)ds%
\Bigg)\Bigg|\exp\Bigg(\frac{\varepsilon \|B\|_{{\sup,T}}^2}{2}\Bigg)\Big|%
{f}_m\Big(z+\sqrt{i}B_{t}\Big)\Big|\Bigg]  \notag \\
\leq& 2\Bigg(\frac{2}{\pi T}\Bigg)^{1/2}\exp\Bigg(T\exp\Bigg(\frac{n}{2}\log%
\Bigg(\frac{x^2}{2}\Bigg)\Bigg)\Bigg)\\&\times\int_0^\infty c_{m,l}(z)u^{m+l}\exp%
\Bigg(\Bigg(\frac{1}{2}+\frac{1}{\sqrt{2}\varepsilon}\Bigg)|z|^2\Bigg)\exp\bigg( \frac{%
\varepsilon}{2}u^2\bigg)e^{-\frac{u^2}{2T}%
}du=:I(z) ,  \label{erwab}
\end{align}
for all $z\in\mathcal{D}$, $0<\varepsilon<\frac{1}{4T}$ and $%
c_{m,l}$ as in Lemma \ref{lemherm}. Again since $c_{m,l}$ is
measurable and locally bounded it follows that the restriction of $I$ to ${\mathcal{O}}$ is also
measurable and locally bounded.
Now we check whether Assumption \ref{ASSII} is true. 
Relation \eqref{abschJ} follows by Lemma \ref{lembm}, Lemma \ref{lemherm} and Lemma \ref{lem1x}.
Again with Lemma \ref{lembm}, Lemma \ref{lemherm} and Lemma \ref{lem1x} and the functions $k_{z,0}$, $k_{z,1}$ and $%
k_{z,2}$ one can show integrability for (\ref{VvonUtt0}) and (\ref{HvonUtt0}), respectively. 
\end{proof}

\begin{kor}
In the same way one can also show the existence of the Feynman
integrand for potentials of the form
\begin{align}
V_0:\mathcal{D}&\to\mathbb{C}  \notag \\
z&\mapsto \frac{a}{(z-b)^n},  \label{1durchh2}
\end{align}
for $a\in\mathbb{C}$, $b\in\mathbb{R}$ and $n\in\mathbb{N}$. Moreover one can
choose linear combination of the potentials given in
(\ref{holpoly}),(\ref{1durchh}) and (\ref{1durchh2}).
\end{kor}
\section*{Acknowledgments}
We would like to thank Michael R\"ockner for valuable discussions. Financial support of Project PTDC/MAT/$67965/2006$ and FCT, POCTI-$219$, FEDER is gratefully acknowledged.
\bibliographystyle{plain}
\bibliography{bibfile}

\begin{thebibliography}{10}

\bibitem{BeKo88}
Yu.M. Berezansky and Yu.G. Kondratiev.
\newblock {\em Spectral Methods in Infinite-Dimensional Analysis}.
\newblock Kluwer Academic Publishers, Dordrecht, 1995.
\newblock Originally in Russian, Naukova Dumka, Kiev, 1988.

\bibitem{BCB02}
Christopher~C. Bernido and M.~Victoria Carpio-Bernido.
\newblock Path integrals for boundaries and topological constraints: a white
  noise functional approach.
\newblock {\em J. Math. Phys.}, 43(4):1728--1736, 2002.

\bibitem{C61}
R.~H. Cameron.
\newblock A family of integrals serving to connect the {W}iener and {F}eynman
  integrals.
\newblock {\em J. Math. Phys.}, 39:126--140, 1960.

\bibitem{FPS91}
M.~de~Faria, J.~Potthoff, and L.~Streit.
\newblock The {F}eynman integrand as a {H}ida distribution.
\newblock {\em J. Math. Phys.}, 32(8):2123--2127, 1991.

\bibitem{FOS05}
Margarida de~Faria, Maria~Jo{\~a}o Oliveira, and Ludwig Streit.
\newblock Feynman integrals for nonsmooth and rapidly growing potentials.
\newblock {\em J. Math. Phys.}, 46(6):063505, 14, 2005.

\bibitem{D80}
Halim Doss.
\newblock Sur une r\'esolution stochastique de l'\'equation de {S}chr\"odinger
  \`a coefficients analytiques.
\newblock {\em Comm. Math. Phys.}, 73(3):247--264, 1980.

\bibitem{GKSS97}
M.~Grothaus, D.C. Khandekar, J.L. Silva, and L.~Streit.
\newblock The {F}eynman integral for time dependent anharmonic oscillators.
\newblock {\em J. Math. Phys.}, 38(6):3278--3299, 1997.

\bibitem{GV08}
M.~Grothaus and A.~Vogel.
\newblock The {F}eynman integrand as a white noise distribution beyond
  perturbation theory.
\newblock To appear in the proceedings of the "5th Jagna International
  Workshop: Stochastic and Quantum Dynamics of Biomolecular Systems", 2008.

\bibitem{Hi80}
T.~Hida.
\newblock {\em {B}rownian Motion}.
\newblock Springer Verlag, Berlin, Heidelberg, New York, 1980.

\bibitem{HKPS93}
T.~Hida, H.-H. Kuo, J.~Potthoff, and L.~Streit.
\newblock {\em White Noise. An infinite dimensional calculus}.
\newblock Kluwer Academic Publisher, Dordrecht, Boston, London, 1993.

\bibitem{KS92}
D.C. Khandekar and L.~Streit.
\newblock Constructing the {F}eynman integrand.
\newblock {\em Ann. Physik}, 1:46--55, 1992.

\bibitem{KLPSW96}
Yu.G. Kondratiev, P.~Leukert, J.~Potthoff, L.~Streit, and W.~Westerkamp.
\newblock Generalized functionals in {G}aussian spaces: The characterization
  theorem revisited.
\newblock {\em J. Funct. Anal.}, 141(2):301--318, 1996.

\bibitem{KWS97}
T.~Kuna, L.~Streit, and W.~Westerkamp.
\newblock Feynman integrals for a class of exponentially growing potentials.
\newblock {\em J. Math. Phys.}, 39(9):4476--4491, 1998.

\bibitem{Kuo96}
H.-H. Kuo.
\newblock {\em White Noise Distribution Theory}.
\newblock CRC Press, Boca Raton, New York, London, Tokyo, 1996.

\bibitem{L06}
Remi Leandre.
\newblock Path integrals in noncommutative geometry.
\newblock In Jean-Pierre Fran{\c{c}}oise, Gregory~L. Naber, and Tsou~Sheung
  Tsun, editors, {\em Encyclopedia of Mathematical Physics (Elsevier, 2006)},
  pages 8--12. Academic Press/Elsevier Science, Oxford, 2006.

\bibitem{P91}
J{\"u}rgen Potthoff.
\newblock Introduction to white noise analysis.
\newblock In {\em Control theory, stochastic analysis and applications
  (Hangzhou, 1991)}, pages 39--58. World Sci. Publ., River Edge, NJ, 1991.

\bibitem{SS02}
Jos{\'e}~L. Silva and Ludwig Streit.
\newblock Feynman integrals and white noise analysis.
\newblock In {\em Stochastic analysis and mathematical physics (SAMP/ANESTOC
  2002)}, pages 285--303. World Sci. Publ., River Edge, NJ, 2004.

\end{thebibliography}

\end{document}